\documentclass{article}
\usepackage{spconf,amsmath, amssymb,hyperref, booktabs,graphicx}
\usepackage{enumitem}
\usepackage{tipa}
\title{Age-Aware Adapter Tuning for Children's Speech Recognition}
\name{Jialu Li}
\address{College of Information Science, University of Arizona}
\begin{document}
\ninept
\maketitle
\begin{abstract}
Children's automatic speech recognition (ASR) remains challenging because child speech differs from adult speech and varies substantially across developmental stages. While adapter tuning provides a promising way to adapt large pretrained ASR models to children’s speech, a single shared child adapter may not fully capture age-dependent variation. In this work, we present one of the first systematic studies of age-aware adapter tuning for child ASR, focusing on speech from children aged 3--12 and older. We propose age-specialized adapters trained separately for different age groups and compare them with a unified age-conditioned FiLM adapter. With ground-truth age routing, age-specialized adapters improve over a strong shared child adapter baseline from 12.5\% to 12.3\% overall
word error rate (WER) and from 16.4\% to 16.1\% macro-age WER,
while consistently improving WER across all four known-age groups.
We further show that predicted-age routing remains close to ground-truth
routing, achieving 12.3\% overall WER and 16.3\% macro-age WER without ground-truth age labels at inference. In contrast, unified FiLM conditioning does not consistently improve over the shared child adapter, indicating that a single unified adapter may be insufficient to capture developmental variation in child speech \footnote{Our code is available at \url{https://github.com/jialuli3/child_asr_age_adapter.git}}.
\end{abstract}
\begin{keywords}
child speech recognition, adapter tuning, parameter-efficient adaptation, age-aware adaptation
\end{keywords}
\vspace{-0.2cm}
\section{Introduction}
\label{sec:intro}

Children's automatic speech recognition (ASR) plays an important role in educational and healthcare technologies, including reading assessment~\cite{harmsen2025can}, pronunciation feedback~\cite{cao2023analysis}, and clinical speech-language applications~\cite{sezgin2022hey}. Despite recent progress in large pretrained ASR models, recognizing children's speech remains challenging. Children's speech data are still scarce in existing speech corpora, and child speech differs substantially from adult speech. Moreover, speech characteristics change rapidly across developmental stages: age-related differences in vocal tract characteristics, pronunciation, speaking style, and linguistic complexity can lead to large performance gaps across age groups~\cite{horii2025children,bhardwaj2022automatic}.

Self-supervised learning (SSL) learns acoustic and linguistic representations from large-scale unlabeled speech, enabling adaptation to downstream ASR tasks with limited labeled data~\cite{yang21c_interspeech, mohamed2022self}. Recent work has shown that fine-tuning or adapting adult-pretrained speech foundation models to children’s speech can improve children’s ASR while reducing reliance on large labeled child-speech corpora and improving robustness to acoustic, phonetic, and developmental variability~\cite{fan24b_interspeech, attia2024kid, liu2024sparsely, jain23_interspeech, ying2025benchmarking, delta_ssl, graave2024mixed, li2026automated}. Among these methods, adapter-based fine-tuning offers a promising parameter-efficient approach for adapting large pretrained speech models to children’s ASR~\cite{fan22d_interspeech,rolland2024exploring}.

A single adapter can effectively adapt a general ASR model to the child-speech domain, but it may not fully capture within-child variability. Prior studies show that ASR performance varies substantially across child age groups, with kindergarten-aged children exhibiting particularly high word error rate (WER)~\cite{osti_10099068}, and that grouping children’s speech by acoustic similarity can improve ASR adaptation~\cite{rolland2025group, Fan2024TowardsBR}. Concurrent work explores a domain-routed Speech-LLM for unified adult and child ASR across multiple corpora, where child age groups are treated as fine-grained domains~\cite{shi2026entropy}. These studies suggest the need for age-aware child ASR studies rather than treating child speech as a homogeneous domain.

In this work, we systematically study whether age information can improve parameter-efficient adaptation of large ASR models for children's speech. We first train a shared child adapter on speech from all children, and then investigate two age-aware adaptation strategies: \textit{age-specialized adapters}, which route separate lightweight adapters to different age groups, and a \textit{unified age-conditioned Feature-wise Linear Modulation (FiLM) adapter}, which conditions a single shared adapter on age information~\cite{perez2018film}. 
%To better assess performance across heterogeneous child speakers, we report overall WER, macro-age-group WER, and group-specific WER.

\textbf{Our contributions are summarized as follows:}
\begin{itemize}[leftmargin=*, itemsep=0pt, topsep=1pt, parsep=0pt]
    \item We present one of the first systematic studies of age-aware adapter tuning for children's ASR, evaluating overall WER, macro-age WER, and age-group WER.
    \item We study two age-aware adaptation strategies, \textit{age-specialized adapters} and a \textit{unified age-conditioned FiLM adapter}, under both ground-truth and router-predicted age settings. We show that age-specialized adapters are more effective than unified FiLM conditioning with ground-truth age information, providing statistically significant gains over a strong shared child adapter baseline in overall and macro-age WER, with improvements across all known-age groups. Predicted-age routing performs close to ground-truth routing, enabling age-aware adapter selection without ground-truth age metadata at inference time.
    \end{itemize}
    %\item We study two age-aware adaptation strategies: \textit{age-specialized adapters} and \textit{a unified age-conditioned FiLM adapter}, under both ground-truth and router-predicted age settings.
    %\item We show that age-specialized adapters are more effective than unified FiLM conditioning with ground-truth age information, providing statistically significant gains over a strong shared child adapter baseline in overall and macro-age WER, with improvements across all known-age groups. Predicted-age routing performs close to ground-truth routing, enabling age-aware adapter selection without ground-truth age metadata at inference time.\end{itemize}

\begin{table}[t]
\centering
\small
\setlength{\tabcolsep}{0pt}
\renewcommand{\arraystretch}{1.15}
\begin{tabular*}{\columnwidth}{@{}l@{\extracolsep{\fill}}cccccc@{}}
\toprule
\textbf{Split} & \textbf{Total} & \textbf{3--4} & \textbf{5--7}
& \textbf{8--11} & \textbf{12+} & \textbf{Unknown} \\
\midrule
Train & 2993 / 288 & 200 / 27 & 1629 / 72 & 1012 / 180 & 148 / 6 & 66 / 3.1 \\
Dev   & 101 / 10.5 & 11 / 1.6 & 52 / 3.0 & 35 / 5.4 & 5 / 0.3 & 4 / 0.2 \\
Test  & 250 / 20.3 & 13 / 1.9 & 133 / 5.2 & 80 / 12.1 & 17 / 0.8 & 11 / 0.4 \\
\bottomrule
\end{tabular*}
\caption{Dataset statistics by age group across train, development, and test splits. Each cell reports the number of speakers / total hours. }
\label{tab:data_stats}
\vspace{-0.4cm}
\end{table}

\vspace{-0.3cm}
\section{Data}
\label{sec:data}
%\vspace{-0.2cm}
We use the Word Track data from the On Top of Pasketti Children's ASR Challenge~\cite{drivendata2026pasketti,bull2016crowd}\footnote{\url{https://www.drivendata.org/competitions/308/childrens-word-asr/page/972/}}, which contains speech from children in the 3–4, 5–7, 8–11, and 12+ (age 12 and older) age groups, along with a subset with unknown age, spanning read speech, classroom and child–robot interactions, narrative language samples, and clinical speech. The data include ReadNet~\cite{readnet}, JIBO Kids~\cite{jibo_kids}, the Arizona Child Acoustic Database Repository~\cite{arizona_child_acoustic_database}, the CMU Kids Corpus~\cite{cmu_kids}, CSLU Kids' Speech~\cite{cslu_kids}, My Science Tutor~\cite{myst}, the Ohio Child Speech Corpus~\cite{ohio_child_speech_corpus}, Cameron~\cite{cameron_project_equity}, the Edmonton Narrative Norms Instrument~\cite{enni}, the Ellis Weismer Corpus~\cite{ellis_weismer}, PERCEPT-GFTA~\cite{ayala2023auditory}, PERCEPT-R~\cite{percept_r}, the Speech Production Repository for Optimizing Use of AI Technologies (SPROUT)~\cite{sprout}, and the TalkBank-hosted challenge corpus~\cite{macwhinney2007talkbank}. We construct speaker-independent train, development, and test splits with no speaker overlap across splits. We use only utterances shorter than 30 seconds for training to fit within GPU memory constraints. Table~\ref{tab:data_stats} summarizes the dataset statistics by age group.
The \textit{3--4}, \textit{5--7}, and \textit{unknown-age} groups contain a mixture of single-word pronunciation samples and short utterances. The \textit{8--11} group primarily consists of longer read-speech utterances, whereas the \textit{12+} group contains only single-word pronunciation samples. 
 %The 8--11 group dominates in duration across all splits, while the 5--7 group contains the largest number of speakers.
\vspace{-0.3cm}

\section{Model}
\begin{figure}[t]
    \centering
    \includegraphics[width=1.0\linewidth]{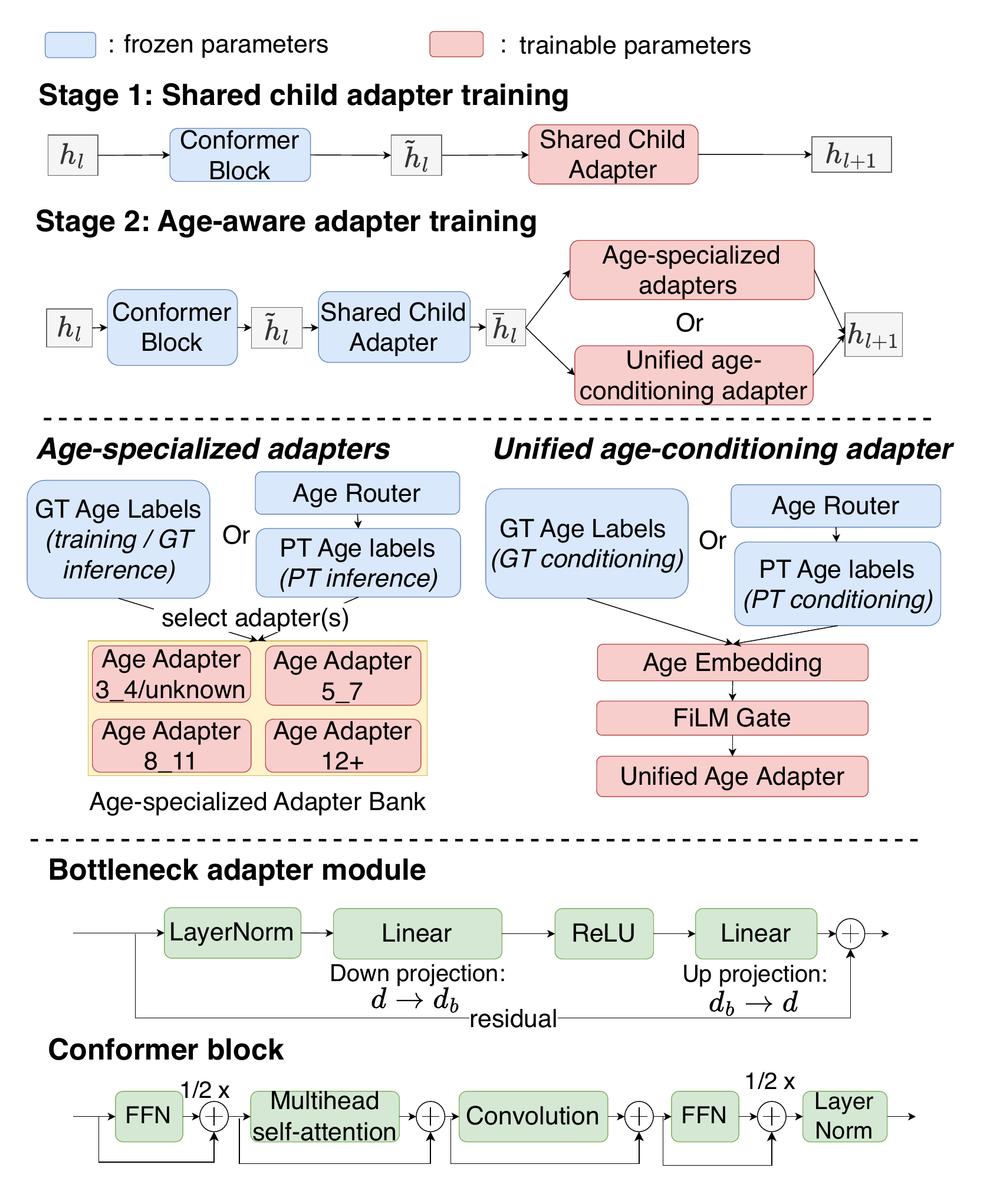}
    \vspace{-0.7cm}

\caption{Overview of the two-stage age-aware adaptation framework. Stage 1 trains a shared child adapter with the ASR backbone frozen; Stage 2 freezes the shared adapter and trains either age-specialized adapters or a unified FiLM adapter. GT/PT denote ground-truth/predicted age generated by age router. Blue modules are frozen and red modules are trainable at each stage. The bottom panels show the bottleneck adapter, which projects from hidden size $d$ to bottleneck size $d_b$ and back, and the Conformer block.}
%\caption{Overview of the proposed two-stage child ASR adaptation framework.
%At layer $l$, the input representation $h_l$ is processed by the frozen Conformer block to obtain $\tilde{h}_l$. In Stage 1, the shared child adapter is trained with the ASR backbone frozen, mapping $\tilde{h}_l$ to $h_{l+1}$. In Stage 2, the trained shared child adapter is frozen and maps $\tilde{h}_l$ to the intermediate representation $\bar{h}_l$, which is further processed by either age-specialized adapters or a unified age-conditioned FiLM adapter to obtain $h_{l+1}$.
%For both age-aware adaptation approaches, we consider two age-label settings: ground-truth (GT) and router-predicted (PT).
%Blue modules are frozen and red modules are trainable at each stage.
%The bottom panels show the bottleneck adapter, which projects from hidden size $d$ to bottleneck size $d_b$ and back, and the Conformer block.}

%\caption{Overview of the proposed child ASR adaptation framework. At layer $l$, the hidden representation $h_l\in\mathbb{R}^d$ is processed by the frozen Conformer block and adapter modules to obtain $h_{l+1}$. A shared child adapter is first trained with the backbone frozen. We then compare age-specialized adapters selected by ground-truth or router-predicted age labels with a unified age-conditioned FiLM adapter. Blue modules are frozen and red modules are trainable; the bottom panels show the bottleneck adapter, which projects from hidden size $d$ to bottleneck size $d_b$ and back, and the Conformer block.}

\vspace{-0.5cm}
\label{fig:model}
\end{figure}

Figure~\ref{fig:model} illustrates the proposed two-stage age-aware adapter framework for child ASR. 
%We first train a shared child adapter on all available data, and then compare two age-aware variants: an age-specialized adapter bank selected by an age router and a unified age adapter modulated by age information through a FiLM gate.

\noindent\textbf{Backbone ASR model.}
We use NVIDIA \textit{Parakeet-tdt-0.6B-v2}\footnote{\url{https://huggingface.co/nvidia/parakeet-tdt-0.6b-v2}} as the pretrained ASR backbone. 
The model is a 600M-parameter English ASR system comprising a 24-layer XL FastConformer encoder~\cite{10389701} with a hidden dimension of 1024 and a Token-and-Duration Transducer (TDT) decoder~\cite{xu2023efficient}. 
%It is initialized from a FastConformer SSL checkpoint pretrained with a wav2vec-style~\cite{baevski2020wav2vec} objective on LibriLight~\cite{librilight}, and subsequently trained on the Granary datasets~\cite{raokoluguri25_interspeech}.
%, which contains approximately 120k hours of English speech, including 10k hours of human-transcribed data and 110k hours of pseudo-labeled data. 
In this study, we freeze the pretrained ASR backbone in all experiments and adapt the 24 encoder layers to child speech using lightweight adapter modules. 

\noindent \textbf{Shared child adapter.}
In Stage 1, we train a shared child adapter on all available child speech data.
For each Conformer layer $l$, the frozen pretrained block produces
\begin{equation}
    \tilde{h}_l = \mathcal{C}_l(h_l;\theta_l),
\end{equation}
where $\mathcal{C}_l$ denotes the pretrained Conformer block, $\theta_l$ denotes its frozen parameters, and $h_l$ is the input hidden representation. We insert a bottleneck adapter after the Conformer block:
\begin{equation}
    h_{l+1}
    =
    \mathcal{A}^{\mathrm{child}}_l(\tilde{h}_l)
    =
    \tilde{h}_l
    +
    W^{\mathrm{up}}_l
    \phi
    \left(
    W^{\mathrm{down}}_l
    \mathrm{LN}(\tilde{h}_l)
    \right),
\end{equation}
where $W^{\mathrm{down}}_l \in \mathbb{R}^{d_b \times d}$ and
$W^{\mathrm{up}}_l \in \mathbb{R}^{d \times d_b}$ are the down- and up-projection matrices, $d=1024$ is the hidden dimension, and $d_b$ is the adapter bottleneck dimension. We denote the bottleneck dimension of the shared child adapter by
$d_b^{\mathrm{child}}$ and set $d_b^{\mathrm{child}}=128$, which keeps the adapter lightweight with about 6.3M trainable parameters. We additionally evaluate $d_b^{\mathrm{child}}=256$ only as a controlled higher-capacity baseline. In Stage 2, we freeze the shared child adapter as the common child-domain module for both age-specialized and unified FiLM training.

\noindent \textbf{Age-specialized adapters.}
For age-aware adaptation, we define four age-based routing groups:
$\mathcal{G}=\{\textit{3\text{--}4/\text{unknown},\allowbreak
5\text{--}7,\allowbreak
8\text{--}11,\allowbreak 12+\}}$.
We merge the \textit{3--4} and \textit{unknown-age} subsets into one routing group, as the \textit{unknown-age} subset lacks a clear age distribution and showed preliminary ASR performance most similar to the \textit{3--4} group. During evaluation, we still report WER for the \textit{unknown-age} subset separately.

\noindent\textbf{\textit{Training.}}
Let $\mathcal{G}$ denote the set of age-group categories used for age-aware adaptation, and let $\mathcal{D}_a$ denote the training subset corresponding to age group $a\in\mathcal{G}$.
The full training set is
$\mathcal{D}_{\mathrm{all}}=\bigcup_{a\in\mathcal{G}}\mathcal{D}_a$.
At layer $l$, the frozen pretrained Conformer block and frozen shared child adapter produce
$\bar{h}_l
    =
    \mathcal{A}^{\mathrm{child}}_l
    \left(
    \mathcal{C}_l(h_l;\theta_l)
    \right)$.
For each age group $a$, we train an age-specialized adapter
$\mathcal{A}^{\mathrm{age}}_{l,a}$ using only utterances from $\mathcal{D}_a$:
\begin{equation}
    h_{l+1}
    =
    \mathcal{A}^{\mathrm{age}}_{l,a}
    \left(
    \bar{h}_l
    \right),
    \qquad x\in\mathcal{D}_a .
\end{equation}

Each age-specialized adapter uses a bottleneck dimension of
$d_b^{\mathrm{age}}=32$, corresponding to approximately 1.6M parameters per adapter and 6.4M parameters across the four age groups.

% \noindent\textbf{\textit{Training.}} Let $\mathcal{D}_a$ denote the training subset for routing group $a\in\mathcal{G}$, and let
% $\mathcal{D}_{\mathrm{all}}=\bigcup_{a\in\mathcal{G}}\mathcal{D}_a$
% denote the full training set. Given the hidden state $h_l$ at layer $l$, the frozen pretrained Conformer block and frozen shared child adapter produce
%     $\bar{h}_l
%     =
%     \mathcal{A}^{\mathrm{child}}_l
%     \left(
%     \mathcal{C}_l(h_l;\theta_l)
%     \right)$.
% For each routing group $a$, we train an age-specialized adapter $\mathcal{A}^{\mathrm{age}}_{l,a}$ using only utterances from $\mathcal{D}_a$:
% \begin{equation}
%     h_{l+1}
%     =
%     \mathcal{A}^{\mathrm{age}}_{l,a}
%     \left(
%     \bar{h}_l
%     \right),
%     \qquad x\in\mathcal{D}_a .
% \end{equation}

% Each age-specialized adapter uses a bottleneck dimension of $d_b^{\mathrm{age}}=32$, corresponding to approximately 1.6M parameters per adapter and 6.4M parameters across four routing groups.

\noindent\textbf{\textit{Inference.}}
We compare two adapter selection settings. In the ground-truth (GT) setting, the GT age group selects the corresponding age-specialized adapter. In the predicted (PT) setting, we first train an age router (described below) and then keep it frozen. Given an input utterance $x$, the router estimates the age posterior $p(a\mid x)$, from which we select the top-$k$ age groups $\mathcal{G}_k\subseteq\mathcal{G}$. For a selected age group $a$, the utterance is passed through the encoder using the same layer-wise computation as in training:
\begin{equation}
h^{(a)}_{l+1}=\mathcal{A}^{\mathrm{age}}_{l,a}
\left(
\mathcal{A}^{\mathrm{child}}_l
\left(
\mathcal{C}_l(h^{(a)}_l;\theta_l)
\right)
\right)
\end{equation}
Let $H^{(a)} = h^{(a)}_L$ denote the final encoder representation for age group $a$, obtained by using the corresponding age-specialized adapter at each adapted layer. For $k>1$, each selected age adapter is evaluated in a separate encoder pass, so each additional selected adapter incurs one additional encoder pass. The resulting encoder representations are combined as
\begin{equation}
H^{\mathrm{top}\text{-}k}
=
\sum_{a\in\mathcal{G}_k}
w_a H^{(a)},
\end{equation}
where $w_a$ is the router posterior weight for age group $a$. Top-1 routing reduces to selecting $\arg\max_{a\in\mathcal{G}} p(a\mid x)$.

\noindent \textbf{Age router.}
The age router is a lightweight two-layer feedforward network with a hidden dimension of 128.
It predicts the age group from the mean-pooled shared-adapter representation at the fourth encoder layer,
$\bar{h}_4=\mathcal{A}^{\mathrm{child}}_4(\mathcal{C}_4(h_4;\theta_4))$:
\begin{equation}
p(a\mid x)
=
\left[
\operatorname{Softmax}
\left(
\mathcal{R}_{\mathrm{ffn}}
\left(
\operatorname{MeanPool}(\bar{h}_4)
\right)
\right)
\right]_a 
\end{equation}
where $\mathcal{R}_{\mathrm{ffn}}(\cdot)$ denotes the feedforward network and
$p(a\mid x)$ is the predicted posterior probability for age group $a$ given input utterance $x$. The age router has around 531K trainable parameters.
We use the fourth-layer shared-adapter representation as input, motivated by prior work showing that age-related acoustic cues in children's speech are prominent in lower encoder layers~\cite{li2024analysis,sinha2026study}.
To avoid selecting the router feature layer based on downstream ASR performance, we fix this feature-extraction layer across all age-aware experiments.
The age router is trained using GT age labels and then frozen for all age-aware experiments.

\noindent\textbf{Unified age-conditioned FiLM adapter.}
As an alternative to separate age-specialized adapters, we train a single
age-conditioned adapter across all age groups. For each utterance,
we construct an age representation $q\in\mathbb{R}^{|\mathcal{G}|}$, which
is either a GT one-hot vector or the posterior distribution
predicted by the frozen age router. This representation is mapped to an
age embedding $e=E^\top q$, where
$E\in\mathbb{R}^{|\mathcal{G}|\times d_e}$ and $d_e=64$. The embedding is
used to modulate the bottleneck representation of the unified adapter
through FiLM conditioning. To keep the model size comparable to the four age-specialized adapters, we set the unified FiLM adapter bottleneck dimension to $d_b^{\mathrm{uni}}=128$, yielding 6.7M trainable parameters. For each layer $l$, we compute
\begin{equation}
\begin{aligned}
%\bar h_l &= \mathcal{A}^{\mathrm{child}}_l(\mathcal{C}_l(h_l;\theta_l)),\\
%z_l &= W^{\mathrm{down}}_l\mathrm{LN}(\bar h_l),\\
\tilde z_l &= z_l+\sigma(g_l)
\left(\gamma_l(e)\odot z_l+\beta_l(e)-z_l\right),\\
h_{l+1} &= \bar h_l+W^{\mathrm{up}}_l\phi(\tilde z_l).
\end{aligned}
\end{equation}
where $z_l = W^{\mathrm{down}}_l\mathrm{LN}(\bar h_l)$ is the adapter bottleneck representation, $\gamma_l(\cdot)$ and $\beta_l(\cdot)$ are learned FiLM scale and shift transformations, and $g_l$ is a learnable gate logit controlling the strength of age conditioning. We initialize the FiLM scale and shift biases to one and zero, respectively, and set $g_l=-3$ so that $\sigma(g_l)\approx 0.05$, keeping the initial module close to the unconditioned adapter. We evaluate both GT and PT conditioning, using the same age source during training and inference. We use age-homogeneous mini-batches so that each update corresponds to a single age condition, and mildly upsample smaller age groups to reduce age imbalance without fully balancing the training distribution.

% \noindent \textbf{Age-homogeneous sampling.}
% For unified FiLM training, we also evaluate age-homogeneous mini-batches,
% where each mini-batch contains utterances from a single age group. To reduce
% age imbalance, we partially upsample smaller groups. Let
% $B_a$ be the original number of mini-batches for age group $a$,
% and $B_{\max}=\max_{b\in\mathcal{G}}B_b$. We set
% \begin{equation}
% B_a^{\mathrm{target}}
% =
% \operatorname{round}\bigl((1-\alpha)B_a
% +\alpha B_{\max}\bigr),
% \end{equation}
% where $\alpha$ controls the upsampling strength. We use $\alpha=0.3$ for mild
% upsampling, which increases minority-group batches without fully balancing the age distribution.

\vspace{-0.2cm}
\section{Experimental Setup}

\noindent\textbf{Training Configuration.} All ASR adapters are trained using the backbone transducer-style loss,
$\mathcal{L}_{\mathrm{ASR}}=-\log p_\theta(y\mid x)$, where $\theta$
denotes the trainable adapter parameters. The age router is trained with
cross-entropy loss, $\mathcal{L}_{\mathrm{age}}=-\log p_\psi(a\mid x)$,
using GT age labels. All models are decoded with the greedy batched TDT decoding configuration.
%, without external language-model rescoring or contextual biasing.
We train the adapter modules using AdamW with batch size 32, learning rate \(1\times10^{-3}\), \(\beta=(0.9,0.98)\), no weight decay, and a cosine annealing schedule with a 0.1 warmup ratio and minimum learning rate \(1\times10^{-5}\). 
Most experiments are trained on a single NVIDIA A100 or A40 GPU. 
For the unified FiLM adapter, we use an RTX PRO 6000 GPU with batch size 64; the FiLM gate and unified age adapter are trained with learning rates \(3\times10^{-4}\) and \(5\times10^{-4}\), respectively. 
%Unless otherwise specified, the shared child adapter is trained for 50k optimization steps. Age-specialized adapters are then trained for 50k steps per age group, and unified FiLM adapters are trained for 60k steps.
We select the checkpoint with the best validation WER for final evaluation.

\noindent\textbf{Compared systems.}
We use the 50K-step shared child adapter as the Stage~1 checkpoint for all two-stage models, and include longer-training (100K steps) and larger-bottleneck ($d_b=256$) shared adapters as controls.
For age-specialized adaptation, we compare adapters (1) trained directly on the frozen ASR backbone alone and (2) on top of the frozen shared child adapter, with GT and PT routing.
For unified age conditioning, we train a FiLM-conditioned adapter on top of the same frozen shared child adapter and evaluate both GT and PT conditioning. As a capacity control, we also train a stacked adapter on the same frozen shared child adapter without age conditioning to determine whether gains arise simply from adding a second adapter layer.

\noindent\textbf{Evaluation.} We evaluate predictions using the official challenge scoring script, which applies the Whisper English normalizer~\cite{radford2023whisper}, and report overall WER, macro-age WER, and age-group WER. Macro-age WER is the unweighted mean over the four known-age groups $\{\textit{3\text{--}4},\allowbreak \textit{5\text{--}7},\allowbreak \textit{8\text{--}11},\allowbreak \textit{12+}\}$. The \textit{unknown-age} subset is reported separately and is included in overall WER but excluded from macro-age WER. %Statistical significance was assessed using a two-sided child-level paired randomization test with 50,000 permutations, with utterances clustered by child \(p<0.05\). For macro age WER, the metric was recomputed within each permutation.
Statistical significance was assessed only for overall WER and macro-age WER using a two-sided child-level paired approximate-randomization test (50,000 permutations; \(p<0.05\)), treating all utterances from the same child as a cluster. For macro-age WER, the four age-specific WERs and their unweighted mean were recomputed within each permutation.

%Significance is assessed against the 100K-step shared child adapter using a two-sided child-level paired randomization test ($p<0.05$).
% \noindent\textbf{Training objectives.}
% All ASR models are trained using the default transducer-style loss of the
% backbone model:
% \begin{equation}
%     \mathcal{L}_{\mathrm{ASR}}
%     =
%     - \log p_\theta(y \mid x),
% \end{equation}
% where \(p_\theta(y\mid x)\) is the probability assigned to the reference
% transcription \(y\) by the adapted ASR model, \(x\) is the input speech, and \(\theta\) denotes the trainable adapter parameters. When
% training the age router with age supervision, we use the cross-entropy loss
% \begin{equation}
%     \mathcal{L}_{\mathrm{age}}
%     =
%     - \log p_\psi(a \mid x),
% \end{equation}
% where \(a\) is the ground-truth age group and \(\psi\) denotes the parameters of
% the age router.

% \noindent \textbf{Evaluation.} 
% We report overall WER, macro WER, and group-specific WER using the official
% challenge scoring script. The unknown-age subset is evaluated separately, and
% macro WER is computed as the unweighted average over
% $\{3\text{--}4,\allowbreak 5\text{--}7,\allowbreak 8\text{--}11,\allowbreak 12+,\allowbreak \mathrm{unknown}\}$.
%Macro WER is the unweighted average over these groups:
% \begin{equation}
%     \mathrm{MacroWER}
%     =
%     \frac{1}{|\mathcal{G}_{\mathrm{eval}}|}
%     \sum_{a \in \mathcal{G}_{\mathrm{eval}}}
%     \mathrm{WER}_a.
% \end{equation}

\begin{table*}[t]
\centering
\setlength{\tabcolsep}{2.0pt}
%\resizebox{\textwidth}{!}{%
\begin{tabular}{l || c c c c c || c | c | c c c c | c}
\toprule
\textbf{Adaptation}
& \shortstack{\textbf{Train}\\\textbf{Age}}
& \shortstack{\textbf{Infer.}\\\textbf{Age}}
& \shortstack{\textbf{Frozen}\\\textbf{Params}}
& \shortstack{\textbf{Trainable}\\\textbf{Params}}
& \shortstack{\textbf{Adapt.}\\\textbf{Steps}}
& \textbf{WER}
& \shortstack{\textbf{Macro-age}\\\textbf{WER}}
& \textbf{3--4}
& \textbf{5--7}
& \textbf{8--11}
& \textbf{12+}
& \textbf{Unk.} \\
\midrule

None/Freeze ASR backbone & -- & -- & -- & -- & --
& 23.8 & 37.1 & 58.2 & 34.2 & 14.7 & 41.1 & 38.8 \\

\midrule
\multicolumn{13}{l}{\textit{Child shared adapter}} \\
\midrule

Child shared ($d_b=128$) & -- & -- & -- & 6.3M & 50k
& 12.6 & 16.6 & 37.8 & 15.1 & 8.5 & 4.9 & 25.5 \\

Child shared ($d_b=128$) & -- & -- & -- & 6.3M & 100k
& 12.5 & 16.4 & 37.4 & 14.9 & 8.5 & 4.8 & 25.8 \\

Child shared ($d_b=256$) & -- & -- & -- & 12.6M & 100k
& 12.5 & 16.6 & 37.4 & \textbf{14.6} & 8.5 & 5.8 & 25.3 \\

\midrule
\multicolumn{13}{l}{\textit{Age-specialized adapters}} \\
\midrule

Age-specialized only & GT & GT & -- & 6.4M & 50k/age
& 12.4 & 16.3 & 37.3 & 14.9 & 8.5 & 4.6 & 24.0 \\

Child shared + age-specialized & GT & GT & 6.3M & 6.4M & 50k + 50k/age
& \textbf{12.3}* & \textbf{16.1}* & \textbf{36.9} & 14.7
& \textbf{8.3} & \textbf{4.3} & \textbf{23.8} \\

Child shared + age-specialized & GT & PT top-1 & 6.8M & 6.4M & 50k + 50k/age
& 12.4 & 16.3 & 37.0 & 14.7 & 8.4 & 5.2 & 24.0 \\

Child shared + age-specialized & GT & PT top-2 & 6.8M & 6.4M & 50k + 50k/age
& \textbf{12.3}* & 16.3 & 37.0 & 14.7
& \textbf{8.3} & 5.2 & \textbf{23.8} \\

\midrule
\multicolumn{13}{l}{\textit{Unified FiLM adapter}} \\
\midrule

Child shared + stacked adapter & -- & -- & 6.3M & 6.3M & 50k + 60k
& 13.1 & 17.2 & 39.5 & 15.5 & 8.7 & 5.1 & 26.7 \\

Child shared + FiLM & GT & GT & 6.3M & 6.7M & 50k + 60k
& 12.7 & 16.5 & 38.1 & 14.9 & 8.6 & 4.5 & 25.4 \\

%Child shared + FiLM, mix & PT all & PT all & 6.8M & 6.7M & 50k + 60k
%& 12.7 & 16.7 & 37.9 & 15.1 & 8.6 & 5.0 & 25.7 \\

Child shared + FiLM & PT all & PT all & 6.8M & 6.7M & 50k + 60k
& 12.6 & 16.3 & 37.0 & 15.0 & 8.5 & 4.7 & 25.3 \\

\bottomrule
\end{tabular}
%}
\caption{WER results (\%) on the test set for the shared child adapter and age-aware adaptation strategies.
\textit{Train Age} and \textit{Infer. Age} indicate the age information used during training and inference, respectively.
\textit{PT top-1/top-2} uses the one or two most probable predicted age adapters.
\textit{PT all} uses the full predicted age posterior.
\textit{Frozen Params} denotes frozen shared child-adapter and, when applicable, age-router parameters.
\textit{Trainable Params} denotes parameters updated in the corresponding adaptation stage.
\textit{Adapt. Steps} denotes the number of optimization steps used for each adaptation stage, reported in order for multi-stage methods.
Best results are shown in \textbf{bold}. * indicates a statistically significant improvement over the shared child adapter baseline ($d_b=128$, 100K steps).}
\label{tab:age_adapter_results}
\vspace{-0.5cm}
\end{table*}

\vspace{-0.2cm}
\section{Results and Discussion}
\label{sec:results}
\subsection{Comparison of age-aware adapters} 
Table~\ref{tab:age_adapter_results} compares different age-aware adaptation strategies. 
Across experiments, the youngest children are the most challenging for ASR, while older children generally show stronger performance, consistent with prior studies~\cite{osti_10099068}. However, the relatively low WER for the \textit{12+} group may partly reflect its simpler single-word pronunciation task. Therefore, rather than interpreting absolute WER differences across age groups as purely developmental effects, we focus on whether each adaptation method improves over the shared child adapter baseline under the same evaluation setting.

\noindent\textbf{Shared child adapter.} Freezing the original ASR model without child-speech adaptation results in substantially higher WER. 
Adding a shared child adapter greatly improves performance, reducing WER to 12.6\% and macro-age WER to 16.6\%. Extending shared child adapter training from 50K to 100K steps yields only marginal gains, while increasing the bottleneck size from 128 to 256 provides no consistent improvement, suggesting that the shared-adapter baseline is near saturation.
%Training the shared child adapter longer for 100K steps or increasing its bottleneck size from 128 to 256 yields only marginal performance improvement, suggesting that the shared-adapter baseline is near saturation.

\noindent\textbf{Age-specialized adapters.} Building on the shared child adapter, age-specialized adapters provide consistent additional gains. With GT routing, age-specialized adapters yield lower overall WER and macro-age WER than the shared child adapter baseline both when used alone and when stacked on top of the frozen shared child adapter, with statistically significant gains in the latter case. The stacked model achieves the best overall WER of 12.3\% and macro WER of 16.1\%, improving all age groups over the shared child adapter baseline ($d_b=128$, steps $=100k$). 
These results suggest that age-specific adaptation provides benefits beyond shared child-domain adaptation.

%PT age routing performs close to GT age routing. Top-1 routing achieves 12.4\% overall WER and 16.3\% macro age WER. Top-2 routing further reduces overall WER to 12.3\%, a statistically significant improvement over the shared child adapter baseline, while maintaining a macro age WER of 16.3\%. Compared with the shared child adapter baseline, top-2 routing improves WER for every age group except \textit{12+}. However, it requires an additional encoder pass and therefore increases inference latency. Since most router errors occur between adjacent age groups (see Figure~\ref{fig:confusion_matrix}), top-2 routing appears to capture most of the useful routing uncertainty. Larger top-$k$ values yield no noticeable additional gains. These results suggest that age-specialized adapter selection can be applied without ground-truth age labels at inference time.

\begin{figure}
    \centering
    \includegraphics[width=0.8\linewidth]{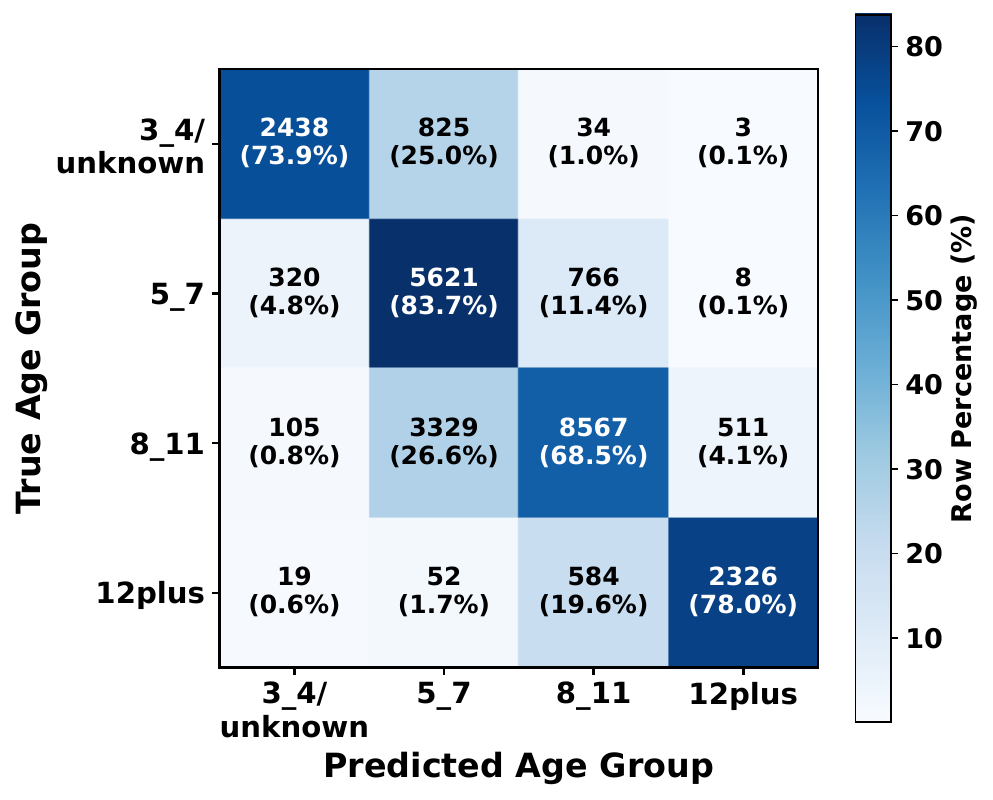}
    \vspace{-0.4cm}
    \caption{Age-router confusion matrix on the test set. Each cell shows the utterance count and row-normalized percentage.}
    \label{fig:confusion_matrix}
    \vspace{-0.5cm}

\end{figure}

PT age routing performs close to GT age routing. Top-1 routing achieves 12.4\% overall WER and 16.3\% macro-age WER, while top-2 routing further reduces overall WER to 12.3\%, a statistically significant improvement over the shared child adapter baseline, with the same 16.3\% macro-age WER. Notably, top-2 routing closely matches GT routing across most age groups and improves over the shared baseline for every age group except \textit{12+}; the remaining degradation relative to GT routing is concentrated in this small \textit{12+} subset, whose data primarily consist of a distinct single-word pronunciation task. Since most router errors occur between adjacent age groups (Figure~\ref{fig:confusion_matrix}), top-2 routing appears to capture most of the useful routing uncertainty. We also evaluated larger top-$k$ values but observed no noticeable additional gains.
Overall, top-2 PT routing offers a favorable trade-off between routing robustness and inference cost, retaining most of the benefit of GT routing with one additional encoder pass.
%whereas top-2 requires an additional encoder pass and therefore increases inference latency. 
%Overall, these results suggest that age-specialized adapter selection can retain most of the benefit of GT routing without requiring GT age labels at inference time.

%\noindent\textbf{Unified FiLM adapter.}
%To separate the effect of age conditioning from simply adding more adapter capacity, we first evaluate a stacked-adapter control that simply adds another shared adapter after the frozen shared child adapter without age conditioning. This control degrades performance from 12.6\% to 13.1\% WER and from 16.6\% to 17.2\% macro WER from child shared adapter baseline, indicating that additional shared capacity alone is not sufficient. Age-conditioned FiLM recovers much of this degradation. With GT age conditioning, the unified FiLM adapter obtains 12.7\% WER and 16.5\% macro WER, while reducing the \textit{12+} WER from 4.9\% to 4.5\%. This suggests that FiLM conditioning can provide benefits for the more homogeneous 12+ single-word pronunciation recordings.
%With PT age conditioning,  we use the full age posterior, which provides a soft age embedding without additional encoder passes and preserves uncertainty across adjacent age groups. Regular age-mixed batching obtains 12.7\% WER and 18.5\% macro WER, whereas age-homogeneous batching improves macro WER to 18.1\% and matches the shared child adapter in overall WER. This suggests that age-homogeneous batching helps the unified adapter make better use of the conditioning signal by reducing within-batch age heterogeneity. 
%However, unified FiLM remains less effective than age-specialized adapters, indicating that soft conditioning in a single shared adapter may not provide sufficient representational separation for developmental variation in child speech.

\noindent\textbf{Unified FiLM adapter.} In contrast, the unified FiLM adapter does not provide the same gains as age-specialized adaptation. GT conditioning achieves 12.7\% overall WER and 16.5\% macro-age WER, while PT conditioning achieves 12.6\% and 16.3\%, respectively. Moreover, the stacked adapter control without age conditioning degrades performance to 13.1\% overall WER and 17.2\% macro-age WER. These results suggest that the gains from age-specialized adapters are not simply due to additional adapter capacity, and that explicitly separating age-specific parameters is more effective than unified FiLM conditioning.

\vspace{-0.3cm}
\subsection{Performance of the age router} 
Figure~\ref{fig:confusion_matrix} shows the confusion matrix of the age router on the test set. 
The router achieves 74.3\% accuracy and 75.7\% macro F1 score. It generally predicts the correct age group, with most errors occurring between adjacent groups. 
%\textit{Unknown-age} test samples are frequently assigned to the merged \textit{3--4/unknown} routing group, consistent with our grouping choice.
%\textit{Unknown-age} samples are frequently routed to the \textit{3--4} group, supporting our decision to merge \textit{unknown-age} samples with the youngest group. 
These results suggest that the router captures useful age-related variation despite ambiguity between neighboring groups, enabling near-GT child ASR performance without perfect age classification.
%These results indicate that the router captures broad age-related acoustic variation, while retaining some ambiguity between neighboring developmental stages. Together with the near-GT ASR performance under PT routing, these results suggest that perfect age classification is not necessary for effective age-aware adapter selection.

%\subsection{Ablation Study of Unified FiLM Adapters}

\vspace{-0.3cm}
\section{Conclusion}
\label{sec:conclusion}
\vspace{-0.2cm}
This work studies whether age information can improve parameter-efficient adaptation for children's ASR. Our results show that age-specialized adapters improve over a strong shared child adapter baseline and outperform unified FiLM conditioning, suggesting that explicitly modeling age-related variation is beneficial for child ASR adaptation. PT routing performs close to GT routing, enabling age-aware adapter selection without GT age metadata at inference time. However, age-group performance should not be interpreted as an effect of chronological age alone: age is associated with developmental differences in utterance length and linguistic complexity and, in our setting, is also confounded with differences in recording and task composition across groups. Overall, these findings support age-specialized residual adaptation as a promising and practical direction for child ASR, while motivating more controlled evaluation that better disentangles age from task and recording differences.

\vspace{-0.2cm}
\section{Acknowledgment}
\vspace{-0.2cm}
This work used the Delta system at the National Center for Supercomputing Applications through allocation CIS260296 from ACCESS program, which is supported by National Science Foundation grants \#2138259, \#2138286, \#2138307, \#2137603, and \#2138296.
\vspace{-0.3cm}

\section{Compliance with Ethical Standards}
\vspace{-0.2cm}
This study used previously collected, de-identified child speech data provided through the DrivenData On Top of Pasketti Children's Speech Recognition Challenge. No new human-subject data were collected for this study.
\vspace{-0.3cm}

{\footnotesize
\bibliographystyle{IEEEbib}
\bibliography{refs}
}

\end{document}